\documentclass[pra,twocolumn,superscriptaddress,amsmath,amssymb,floatfix]{revtex4-2} 
\usepackage{graphicx} 
\usepackage{bm}       
\usepackage[colorlinks=true,linkcolor=blue,urlcolor=blue,citecolor=blue,linkcolor=blue,pdfstartview=FitH]{hyperref}

\usepackage{enumerate}
\usepackage{amsmath}
\usepackage{xcolor}
\usepackage{soul}
\usepackage{cases}
\usepackage{microtype}
\usepackage{siunitx}
\usepackage{subdepth}
\usepackage[utf8]{inputenc}
\usepackage[T1]{fontenc}
\usepackage{txfonts} 

\begin{document}

\title{Active Soft-Impact Oscillator: Dynamics of a Walking Droplet in a Non-Smooth Potential}
\author{Titir Mukherjee}
\affiliation{Department of Physical Sciences, Indian Institute of Science
Education and Research Kolkata, Nadia 741 246, West Bengal, India}
\author{Rahil N. Valani}
\affiliation{Rudolf Peierls Centre for Theoretical Physics, Parks Road,
University of Oxford, OX1 3PU, United Kingdom}
\author{Soumitro Banerjee}
\affiliation{Department of Physical Sciences, Indian Institute of Science
Education and Research Kolkata, Nadia 741 246, West Bengal, India}

\begin{abstract}
Walking droplets are millimetric fluid drops that propel themselves across a vibrated liquid bath through interaction with their self-generated waves. They constitute classical active wave-particle entities and exhibit a range of hydrodynamic quantum analogs. We investigate an \emph{active soft-impact oscillator} as a minimal model for a walking droplet moving within a piecewise-smooth external potential, analogous to classical mass-spring soft-impact oscillators and recently explored quantum soft-impact oscillators. Our active soft-impact oscillator model couples a non-smooth soft-impact force to the Lorenz-like dynamics arising from the wave-particle entity. Theoretical and numerical exploration of the full parameter space reveals a wide variety of nonlinear behaviors and bifurcations driven by impact and grazing events. These include grazing-induced and impact-induced transitions between periodic and chaotic motion, as well as grazing-mediated attractor switching and impact-free (invisible) attractor switching. The active soft-impact oscillator thus provides a versatile platform for probing nonlinear impact dynamics in active systems and exploring hydrodynamic quantum analogs in non-smooth potentials.

\end{abstract}

\maketitle
\section{Introduction}


In recent years, a hydrodynamic system of millimeter-sized walking droplets~\citep{Couder2005, Couder2005WalkingDroplets} has demonstrated that classical systems with memory effects can exhibit several hydrodynamic quantum analogs (HQAs)~\citep{Bush2020review}. In this system, a droplet bounces periodically while walking horizontally on the free surface of a vertically vibrating bath of the same liquid. Each bounce generates a localized and slowly decaying standing wave on the surface, and the droplet subsequently interacts with these self-generated waves. This feedback loop gives rise to self-propelled walking and superwalking droplets~\citep {superwalker}. 

Three key features characterize this hydrodynamic system: (i) the droplet and its wave field coexist as a coupled \emph{wave–particle entity} (WPE); without the droplet the waves decay, and without the waves the droplet ceases to walk, (ii) the droplet is an \emph{inertial active particle}, since it extracts energy locally from the vibrating bath and converts it into directed motion, and (iii) the system possesses \emph{memory}, as past waves influence future motion, embedding history-dependence into the walking dynamics. In this sense, walking droplets may be regarded as examples of \emph{inertial active particles with memory}, placing them in the broader context of active matter~\citep{Ramaswamy_2017, 10.1063/1.5134455}. 


A wide range of HQAs have been identified for walking droplets in experiments. Confinement in central force potentials, for instance, produces discrete sets of quantized orbits such as circles, ovals, lemniscates, and trefoils~\citep{Perrard2014a, Labousse_2014,labousseharmonic, Kurianskiharmonic,durey2018, Perrard2018}. Droplets encountering submerged barriers can exhibit tunneling-like behavior by unpredictably crossing the barrier despite being reflected most of the time~\citep{Eddi2009}. In confined geometries or under external potentials, intermittent chaotic motion can emerge, leading to coherent wave-like statistics reminiscent of those found in quantum systems~\citep{PhysRevE.88.011001,durey_milewski_wang_2020, Giletcircularcavity, Giletcircularcavity2}. For a comprehensive overview of HQAs, we refer the reader to Ref.~\citep{Bush2015, Bush2020review}.

Alongside these experimental discoveries, a series of theoretical models have been developed to capture walker dynamics~\citep{Turton2018, Rahman2020review}. The stroboscopic integro-differential model of Oza \textit{et al.}~\citep{Oza2013} provides a tractable description of horizontal walking dynamics, and has successfully reproduced many HQAs~\citep{Oza2014,harris_bush_2014,labousse2016b, Spinstates, Kurianskiharmonic, Tambascoorbit, ValaniHOM, Saenz2021}. Building on this, the framework of generalized pilot-wave dynamics~\citep{Bush2015,  Bush2020review} has enabled the exploration of WPEs in simplified settings, including one-dimensional models where the integro-differential system reduces to Lorenz-like dynamical equations~\citep{phdthesismolacek, Durey2020, Durey2020lorenz, ValaniUnsteady, Valanilorenz2022}. These Lorenz-like models have revealed nonlinear dynamical mechanisms underlying several HQA setups such as anomalous transport in linear potentials~\citep{Valani2022ANM, Valani2024ALC}, megastable quantization in harmonic potentials~\citep{Lopezvalanimegastablity}, interference effects in sinusoidal potentials~\citep{Perks2023}, tunneling across a barrier~\citep{ZuValani2025} and in a double well potential~\citep{VALANI2024115253}, and emergence of Friedel-like oscillations~\citep{2512.21049}.


In parallel, there is a long tradition of studying \emph{impact oscillators} in classical dynamics as canonical model systems for investigating nonlinear phenomena, motivated in part by their ubiquity in engineering applications. An impact oscillator typically consists of a harmonic oscillator interacting with rigid or soft boundaries. In the absence of impact, the particle evolves under a smooth harmonic potential, while at impact events, its velocity is modified by the boundary interaction. Classical studies have shown that such piecewise-smooth systems generically exhibit grazing bifurcations, border-collision phenomena, multistability, and abrupt transition to chaos \citep{Nordmark1991, Chin1994, Banerjee2009, diBernardo2008}. Quantum analogs of impact oscillators have also been explored \citep{Acharya2023SNC, Acharya2025QuantumImpactSSRN, Mukherjee2025SoftImpactQLE}. Together, these works highlight that non-smooth potentials constitute a particularly rich setting for the emergence of non-trivial bifurcations and complex dynamical behavior in both classical and quantum systems, motivating their extension to active particles and, in particular, walking droplets that have been shown to exhibit several hydrodynamic quantum analogs.


In this paper, we investigate the dynamics of a one-dimensional wave--particle entity (WPE) in a \emph{non-smooth soft-impact potential} using a minimal Lorenz-like model. By coupling memory-driven WPE dynamics to a piecewise-smooth confining potential, we uncover dynamical behavior that is qualitatively distinct from that of classical impact oscillators. In particular, the nonlinear interaction between internal wave-memory degrees of freedom and non-smooth confinement gives rise to extended regimes of both weak and strong chaos, which are absent in traditional impact systems where chaos typically emerges abruptly at grazing. We further show that in the presence of multistability, non-smooth boundaries can induce attractor switching without direct impacts. Together, these results establish the active soft-impact oscillator as a minimal framework in which activity, memory, and non-smooth confinement jointly generate rich and novel nonlinear dynamics.

The paper is organized as follows. In Sec .~\ref {Sec: theory} we present the Lorenz-like dynamical system governing WPE dynamics in a non-smooth potential. We then find equilibrium states of the system and perform a linear stability analysis in Sec.~\ref{analysis}, followed by a numerical exploration of the parameter space dynamics and bifurcations in Sec.~\ref{results}. We discuss and conclude in Sec .~\ref {Sec: Conc}.

\section{Theoretical Model}\label{Sec: theory}

\begin{figure}[!h]
\centering
\includegraphics[width=\columnwidth]{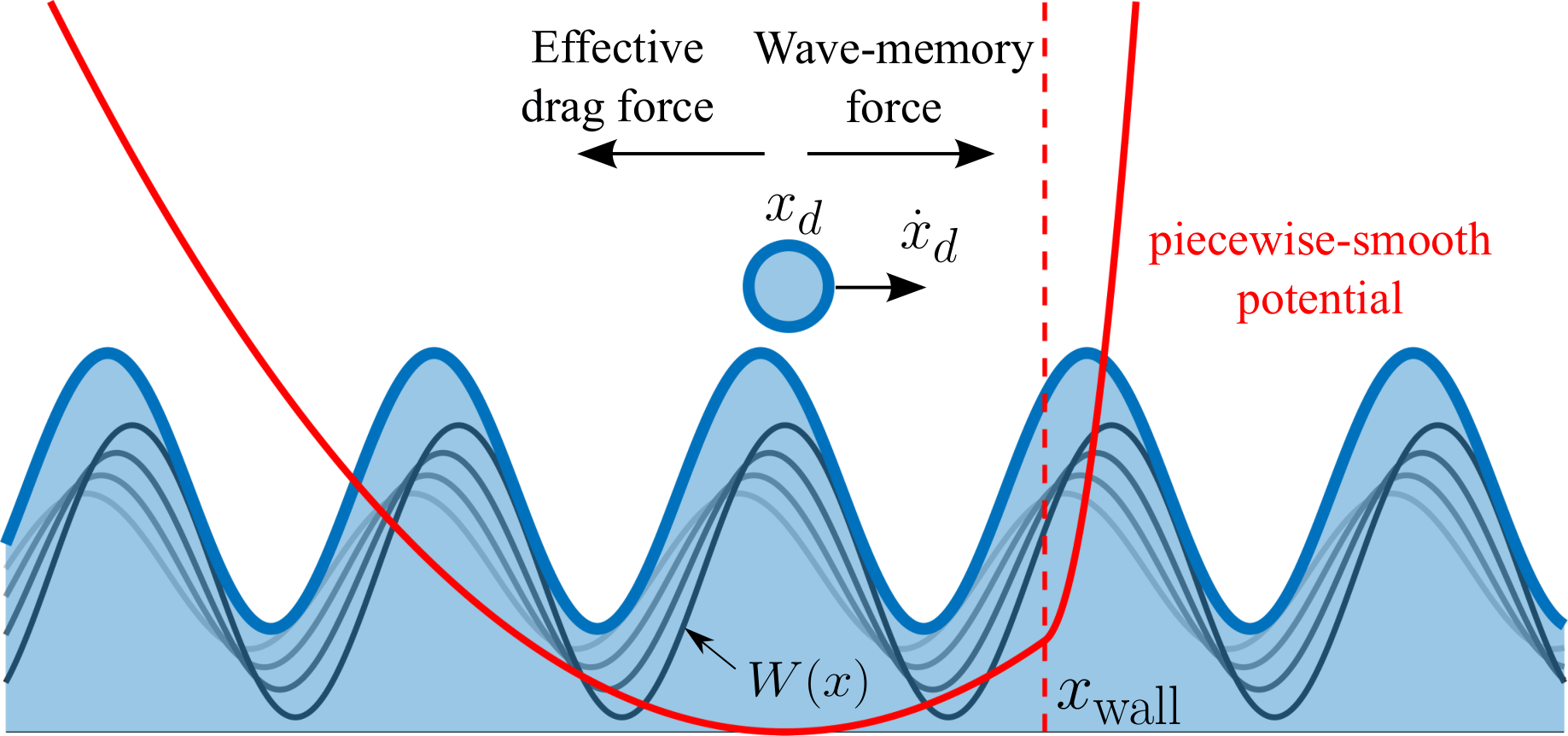}
\caption{Schematic of the setup showing a one-dimensional active wave-particle entity (blue) in a non-smooth harmonic potential (red). A particle (blue circle) of unit dimensionless mass located at $x_d$ and moving horizontally with velocity $\dot{x}_d$ experiences a wave-memory force from its self-generated wave field (blue filled area), and an effective drag force. The underlying wave field is a superposition of the individual waves of spatial form $W(x)=\cos(x)$, which are generated by the particle continuously along its trajectory and decay exponentially in time (black and gray curves). An external piecewise-smooth harmonic potential exerts an additional force on the particle.}
\label{Fig1 schematic}
\end{figure}

As shown schematically in Fig.~\ref{Fig1 schematic}, consider a particle (droplet) at position $x_d$ moving horizontally with velocity $\dot{x}_d$, continuously generating standing waves with prescribed spatial structure $W(x)$ that decay exponentially in time. The particle experiences an external non-smooth "soft-impact" potential of the form
\begin{equation}
V(x) = 
\begin{cases}
\frac{1}{2}\bar{k} x^2, & x < x_\text{wall}, \\
\frac{1}{2}\bar{k} x^2 + \frac{1}{2}\bar{k} {A} (x-x_\text{wall})^2, & x \ge x_\text{wall},
\end{cases}
\end{equation}
where $x_\text{wall}$ is the location of the nonsmoothness i.e. the ``wall" and $\bar{k}$ is a spring constant and ${A}$ is a stiffness parameter. The corresponding force on the droplet is
\begin{equation}
F(x_d) = -\frac{dV}{dx}\Big|_{x=x_d} =
\begin{cases}
-\bar{k} x_d, & x_d < x_\text{wall}, \\
-\bar{k} x_d-\bar{k} {A}(x_d-x_\text{wall}), & x_d \ge x_\text{wall}.
\end{cases}
\end{equation}

The dimensional equation of motion for the horizontal droplet dynamics is~\citep{Oza2013}
\begin{align}\label{Eq: dimensional eq_impact}
&m \ddot{x}_d + D \dot{x}_d = \\ \nonumber
&-\frac{F_m}{T_F} \int_{-\infty}^t W'(k_F(x_d(t)-x_d(s)))\, \text{e}^{-(t-s)/(T_F \text{Me})}\, \text{d}s
+ F(x_d),
\end{align}
where\\
\begin{tabular}{lp{3.1in}}
$m$ & is the droplet mass, \\$D$ &is the effective time-averaged drag coefficient,\\ $k_F$ & $=2\pi/\lambda_F$  is the Faraday wavenumber with $\lambda_F$ the Faraday wavelength, \\
$F_m$ & $=m g A_m k_F$ is a non-negative wave-memory force coefficient with $g$ as the gravitational acceleration and $A_m$ the amplitude of surface waves, \\
$\text{Me}$ & is the memory parameter that describes the proximity to the Faraday instability\\
$T_F$ & is the Faraday period, i.e. the period of droplet-generated standing waves and also the bouncing period of the walking droplet.
\end{tabular}

The left-hand side of Eq.~\eqref{Eq: dimensional eq_impact} is composed of an inertial term $m \ddot{x}_d$ and an effective drag term $D \dot{x}_d$, where the overdot denotes a time derivative. The first term on the right-hand side of the equation captures the time-delayed self-force on the droplet mediated by its underlying wave field. This force is proportional to the gradient of the underlying wave field. The wave field is calculated through the integration of the individual waveforms $W(x)$ that are continuously generated by the particle along its trajectory and decay exponentially in time ($W^\prime$ is the derivative of $W$ with respect to $x$).
 We refer the interested reader to Ref.~\cite{Oza2013} for more details and explicit expressions for the parameters. 

Non-dimensionalizing with $t'=Dt/m$ and $x'=k_F x$, and dropping primes gives
\begin{align}\label{eq: dimless eq2_impact}
\ddot{x}_d + \dot{x}_d = 
-R \int_{-\infty}^t W'(x_d(t)-x_d(s))\, \text{e}^{-(t-s)/\tau}\, \text{d}s + F(x_d),
\end{align}
with $R = m^3 g A_m k_F^2/(D^3 T_F)$ the dimensionless wave amplitude and $M = D T_F \text{Me}/m$ the dimensionless memory. The dimensionless external force is given by
\begin{equation}
F(x_d)  =
\begin{cases}
-{k} x_d, & x_d < x_\text{wall}, \\
-{k} x_d-{k} {A}(x_d-x_\text{wall}), & x_d \ge x_\text{wall}.
\end{cases}
\end{equation}
where $k=m\bar{k}/D^2$ is the dimensionless spring constant and $x_\text{wall}$ is the dimensionless wall position, which we restrict to $x_\text{wall}>0$ in the present study.

In experiments, the spatial form of the individual waves generated by a WPE is captured reasonably well by $W(x)=\text{J}_0(x)$, where $\text{J}_0(x)$ is the Bessel function of the first kind and the zeroth order~\citep{Oza2013, Molacek2013DropsTheory}. Sometimes, a spatially decaying exponential envelope is included to further improve the comparison with the experimentally observed waveform~\citep {Turton2018}. There are two key features of the droplet-generated individual waves: (i) spatial oscillations and (ii) spatial decay. \citet{Durey2020} and \citet{ValaniUnsteady} show that oscillations play a key role in capturing the instability of the steady walking state, and such an instability can be qualitatively captured using a simple sinusoidal particle-generated waveform such that $W(x)=\cos(x)$. The principal advantage of this simple waveform is that it allows us to transform the integrodifferential equation to a low-dimensional Lorenz-like system~\cite{Durey2020lorenz, Valanilorenz2022} by allowing us to define
\begin{align*}
Y(t) &= R \int_{-\infty}^{t} \sin(x_d(t)-x_d(s))\, \text{e}^{-(t-s)/\tau}\, \text{d}s,\\
Z(t) &= R \int_{-\infty}^{t} \cos(x_d(t)-x_d(s))\, \text{e}^{-(t-s)/\tau}\, \text{d}s.
\end{align*}
Differentiating $Y(t)$ and $Z(t)$ with respect to time, we obtain the Lorenz-like four-dimensional ODE system for the WPE dynamics:
\begin{eqnarray}
\dot{x}_d &=& X, \nonumber\\
\dot{X} &=& Y - X + F(x_d),\nonumber\\
\dot{Y} &=& -\frac{1}{M}Y + X Z, \nonumber \\
\dot{Z} &=& R - \frac{1}{M}Z - X Y,
\label{Eq: Lorenz eq_impact}
\end{eqnarray}
where $X=\dot{x}_d$ is the velocity of the particle, $Y$ is the wave-memory force on the particle, $Z$ is the height of the wave-field at the location of the particle, and $F(x_d)$ is the external force of soft-impact.  
This formulation enables the study of a WPE under a \textit{non-smooth, piecewise harmonic potential}. Equations~\eqref{Eq: Lorenz eq_impact} may also be viewed as an impact oscillator for an active particle with internal degrees of freedom. Unlike a classical mass-spring impact oscillator that is driven externally, in our system the active particle is subject to an additional force arising from its internal state, characterized by the variables $Y$ and $Z$. For details of numerical implementation, see Appendix \ref{App: num details}.

\section{Equilibrium point and Linear Stability analysis} \label{analysis}

We start by finding equilibrium states of the dynamical system in Eq.~\eqref{Eq: Lorenz eq_impact}. For $x_{\text{wall}}>0$, this results in a single equilibrium point given by
\begin{equation}
    (x^*_{d}, X^*, Y^*, Z^*) = (0, 0, 0, MR).
    \label{eq: fixed_pt}
\end{equation}

This fixed point corresponds to a stationary wave–particle entity (WPE) resting at the minimum of the external harmonic potential well. To assess the stability of this equilibrium, we carry out a linear stability analysis. Because the equilibrium lies entirely within the region $x_{\mathrm{d}} \leq x_{\mathrm{wall}}$, the force derivative is simply $F'(x_{\mathrm{d}}) = -k$, reflecting the linear restoring nature of the potential in this domain. Evaluating the Jacobian about the equilibrium point~\eqref{eq: fixed_pt}, we obtain
\begin{equation}
J_{\mathrm{ep}} =
\begin{pmatrix}
0 & 1 & 0 & 0 \\
-k & -1 & 1 & 0 \\
0 & MR & -\tfrac{1}{M} & 0 \\
0 & 0 & 0 & -\tfrac{1}{M}
\end{pmatrix},
\label{eq: Jacobian}
\end{equation}
which governs the linearized dynamics in the vicinity of the stationary state.

\begin{figure*}[hbtp]
    \centering
    \includegraphics[width=0.3\textwidth]{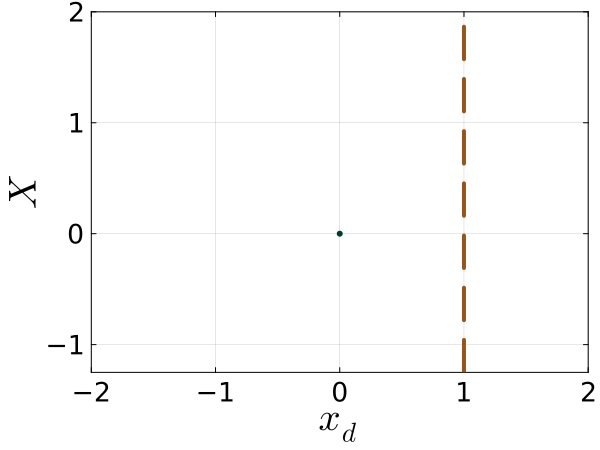}
    \includegraphics[width=0.3\textwidth]{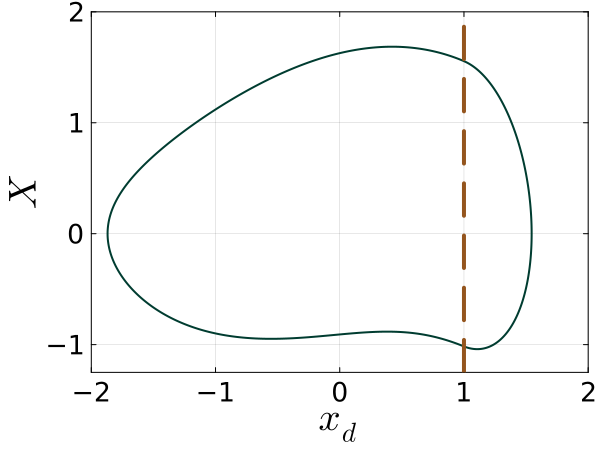}
    \includegraphics[width=0.3\textwidth]{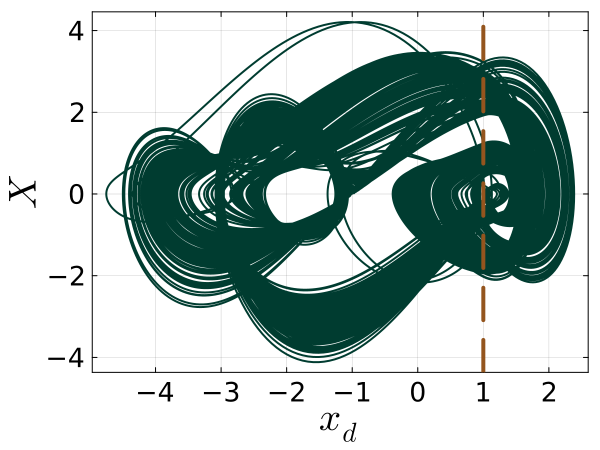} \\
    {\footnotesize (a)\hspace{4.5cm}(b)\hspace{4.5cm}(c)}
    \caption{{\color{black}Different types of WPE dynamical behaviors observed in the system with increasing memory parameter $M$. Phase-space trajectory in the $(x_d,X)$ projection showing (a) $M = 0.5$, stable fixed point, (b) $M = 2$, a stable periodic orbit, (c) $M=10$, a chaotic attractor. The vertical dashed line denotes $x=x_{\text{wall}}$. Other parameters are fixed to $k=1$, $A=5$, $x_{\text{wall}}=1.0$ and $R=1.5$ and initial condition $(x_d(0),X(0),Y(0),Z(0))=(1.0,0.1,0,0)$.}}
    \label{fig: trajs}
\end{figure*}

The stability of this stationary state is governed by the eigenvalues $\lambda$ of the Jacobian,  which satisfy the characteristic equation.
\[
\det(J_{\text{ep}} - \lambda I) = 0.
\]

This yields
\begin{equation}
\left(\lambda + \frac{1}{M}\right)
\left[
\lambda^{3}
+ \left(1 + \frac{1}{M}\right)\lambda^{2}
+ \left(\frac{1}{M} - MR + k\right)\lambda
+ \frac{k}{M}
\right]
= 0.
\end{equation}

One of the eigenvalues is $\lambda = -\tfrac{1}{M} < 0$ (as $M > 0$). The remaining three eigenvalues are determined by the cubic polynomial inside the brackets. The fixed point loses stability when any of these eigenvalues acquires a positive real part. The Routh--Hurwitz criterion provides a systematic method to determine whether all roots of a characteristic polynomial possess negative real parts without explicitly solving for them~\cite{routh1877stability,hurwitz1895stability,ogata2010modern}. This leads to the analytical expression for the stability boundary for our system:
\begin{equation}
    R = \frac{1}{M^2}+ \frac{k}{M + 1}.
    \label{eq: stability_curve}
\end{equation}

Hence, the equilibrium state $(x^*_{d}, X^*, Y^*, Z^*) =(0, 0, 0, MR)$ remains stable when the parameters $( R, M)$ lie below this curve. As the system crosses this boundary, a Hopf bifurcation occurs: the stable fixed point becomes unstable, and gives rise to a limit cycle.

\begin{figure*}[hbtp!]
    \centering
    \includegraphics[width=0.31\textwidth]{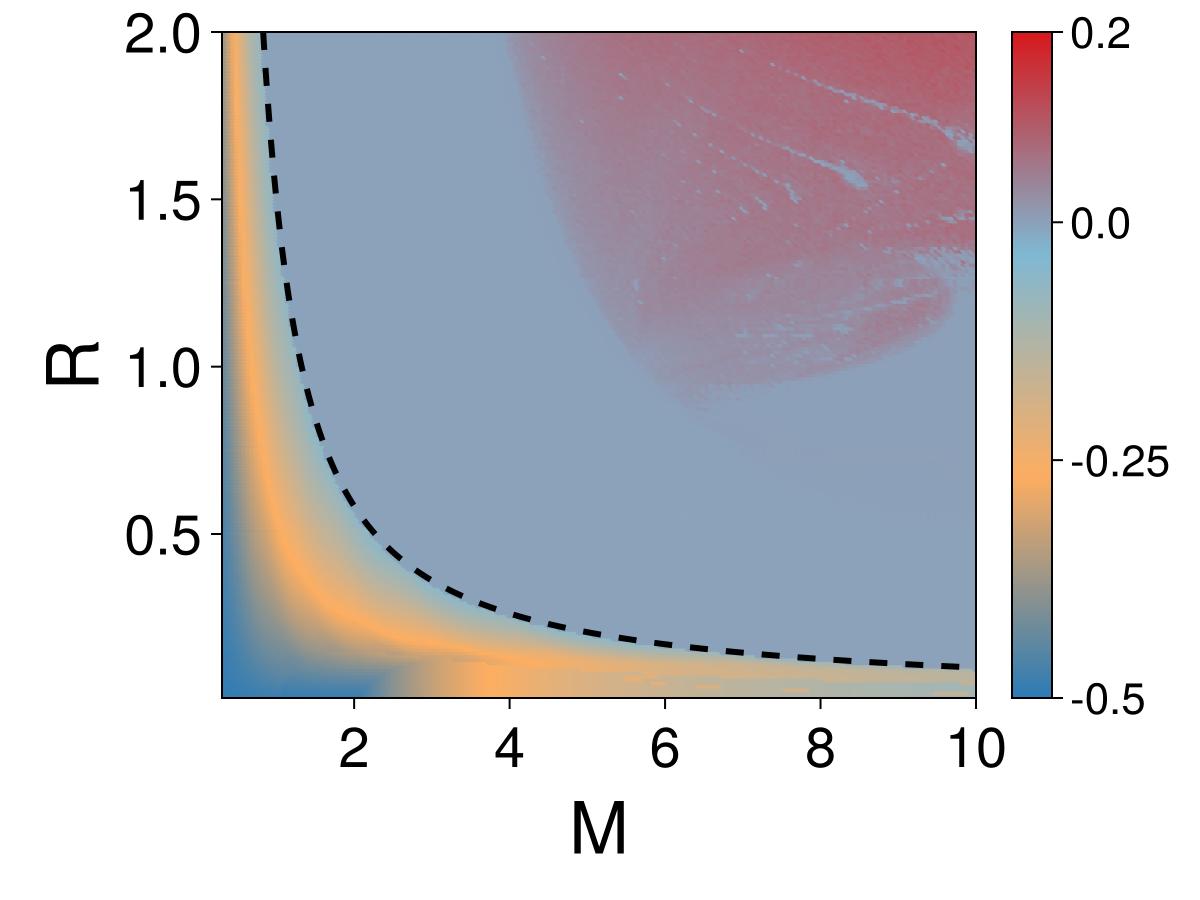}
    \includegraphics[width=0.31\textwidth]{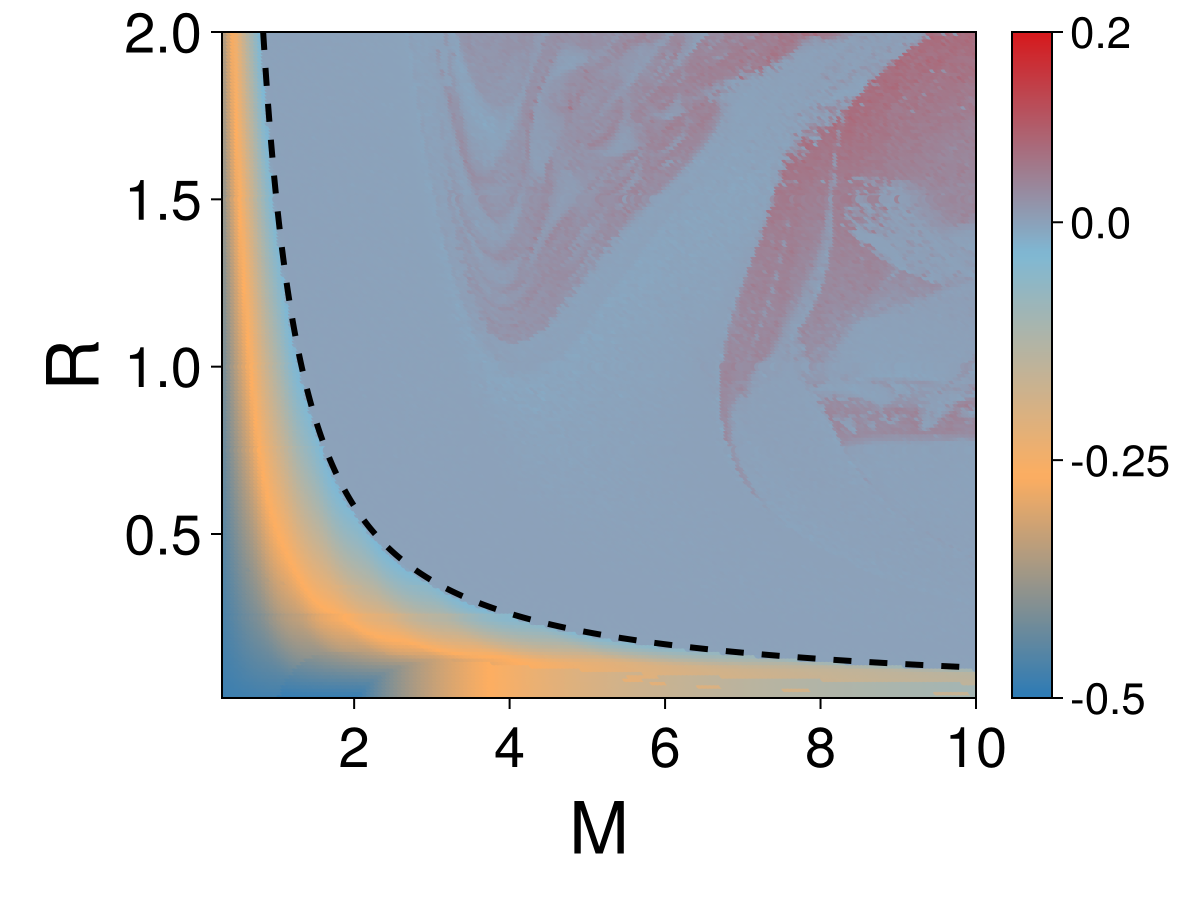}
    \includegraphics[width=0.31\textwidth]{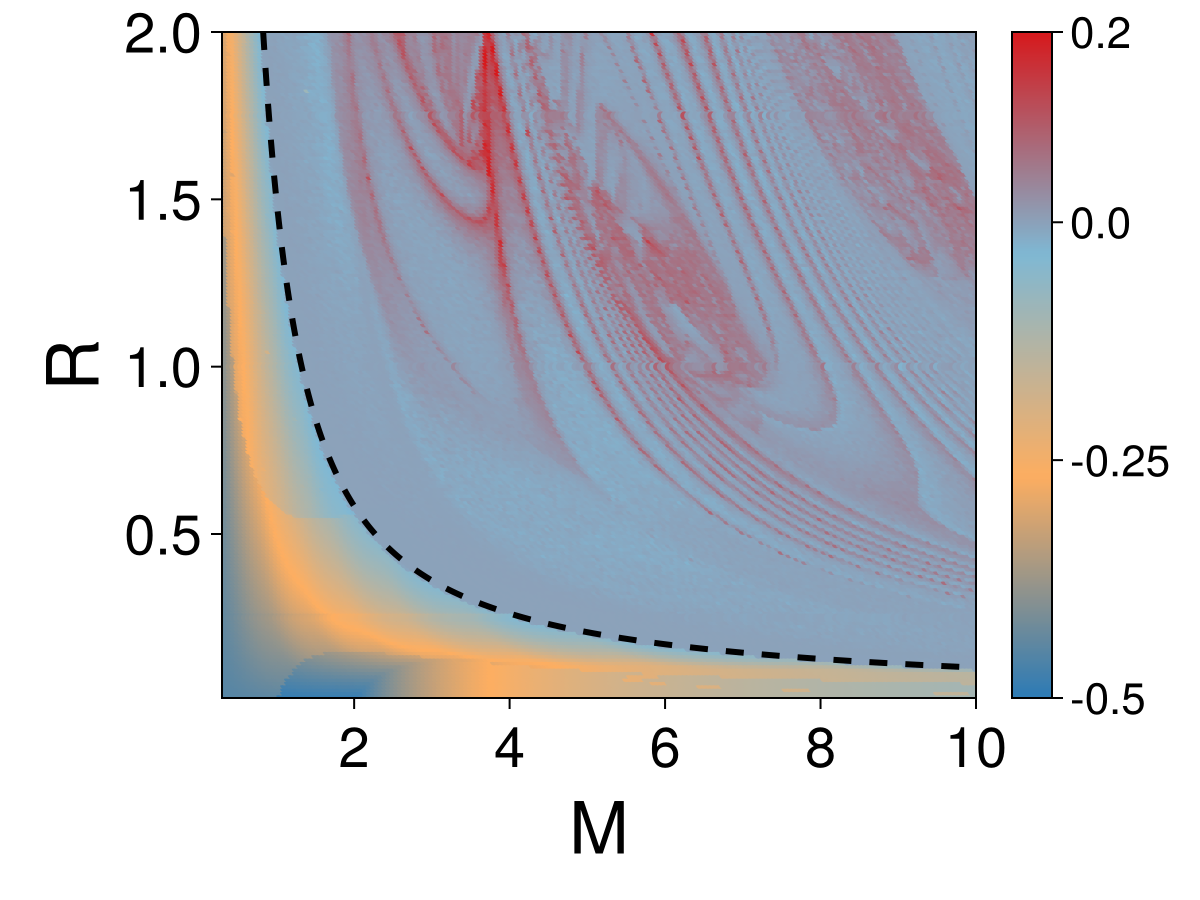}\\
    {\footnotesize (a)\hspace{5cm}(b)\hspace{5cm}(c)}\\[2mm]
    \caption{Dynamics in the $R$-- $M$ parameter space for three values of the stiffness parameter $A$, where the color denotes the maximum Lyapunov exponent (MLE). The black dashed curve indicates the analytical stability boundary given by Eq.~\eqref{eq: stability_curve}. Regions with negative, zero, and positive MLE correspond to fixed-point, periodic/quasiperiodic dynamics, and chaotic dynamics, respectively: (a) $A = 0$, (b) $A = 5$, and (c) $A = 100$. Other parameters are fixed to $k=1$, $x_{\text{wall}}=1.0$, and initial conditions are fixed to $(x_d(0),X(0),Y(0),Z(0))=(1.0,0.1,0,0)$.}

\label{fig: RM}
\end{figure*}

\section{Parameter-space exploration} \label{results}

In this section, we present the results of our numerical exploration of the dynamics and bifurcations observed in the motion of the WPE in the impact oscillator potential, with variations in the dimensionless memory parameter $M$, dimensionless wave amplitude $R$, wall position $x_\text{wall}$ and stiffness parameter $A$.

\subsection{Dynamics in the $R$--$M$ parameter space}\label{R-M}

We begin by presenting the qualitatively distinct types of WPE motion in the phase plane formed by the position $x_d$ and the velocity $X$, as illustrated in Fig.~\ref{fig: trajs}. In these graphs, the memory parameter $M$ is progressively increased while all other parameters are fixed at $k = 1$, $R = 1.5$, $A = 5$, and $x_{\text{wall}} = 1.0$. For a small memory value $M = 0.5$, shown in Fig.~\ref{fig: trajs}(a), the WPE relaxes to a stationary state at $x_d = 0$. In this regime, the wave field generated by the particle decays rapidly, so the memory-induced propulsion is weak and insufficient to overcome the dissipation and the restoring force of the confining potential. As a result, the particle settles at the minimum of the potential and remains trapped there.

As the memory parameter increases to $M = 2$, a sustained oscillatory motion emerges, as shown in Fig.~\ref{fig: trajs}(b). In this intermediate-memory regime, the wave field persists long enough for the WPE to interact with its own past trajectory, leading to a self-consistent balance between wave-induced propulsion, effective drag, and the restoring force of the potential. This balance gives rise to a stable limit cycle in phase space, corresponding to the periodic back-and-forth motion of the particle within the confining potential. 

For even larger memory, $M = 10$, the WPE dynamics becomes chaotic and irregular, as illustrated in Fig.~\ref{fig: trajs}(c). Here, the long-lived wave field leads to strong feedback that destabilizes the periodic orbit, resulting in irregular excursions in phase space. The sensitive dependence on initial conditions is indicated by a positive Lyapunov exponent $\approx 0.04$.

\begin{figure*}[hbtp!]
    \centering
    \includegraphics[width=0.31\textwidth]{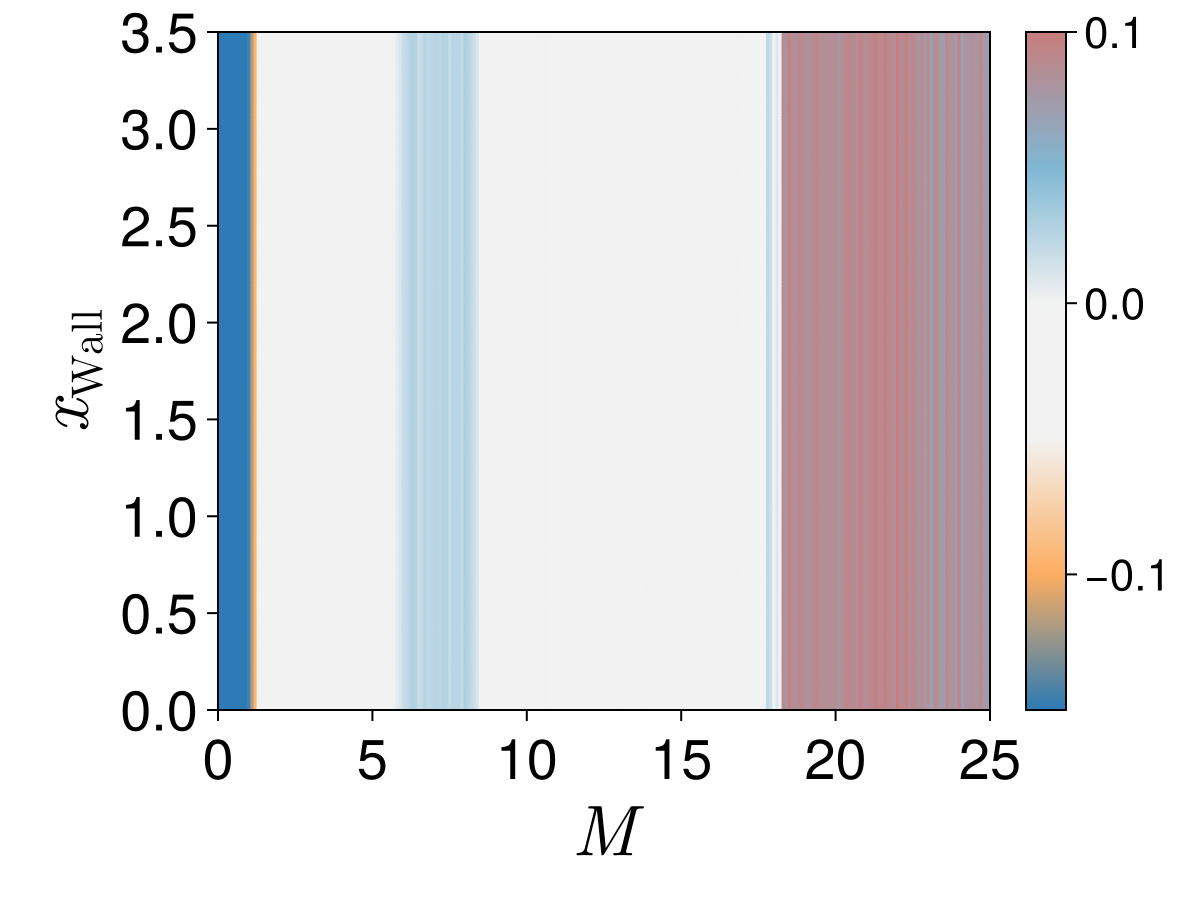}
    \includegraphics[width=0.31\textwidth]{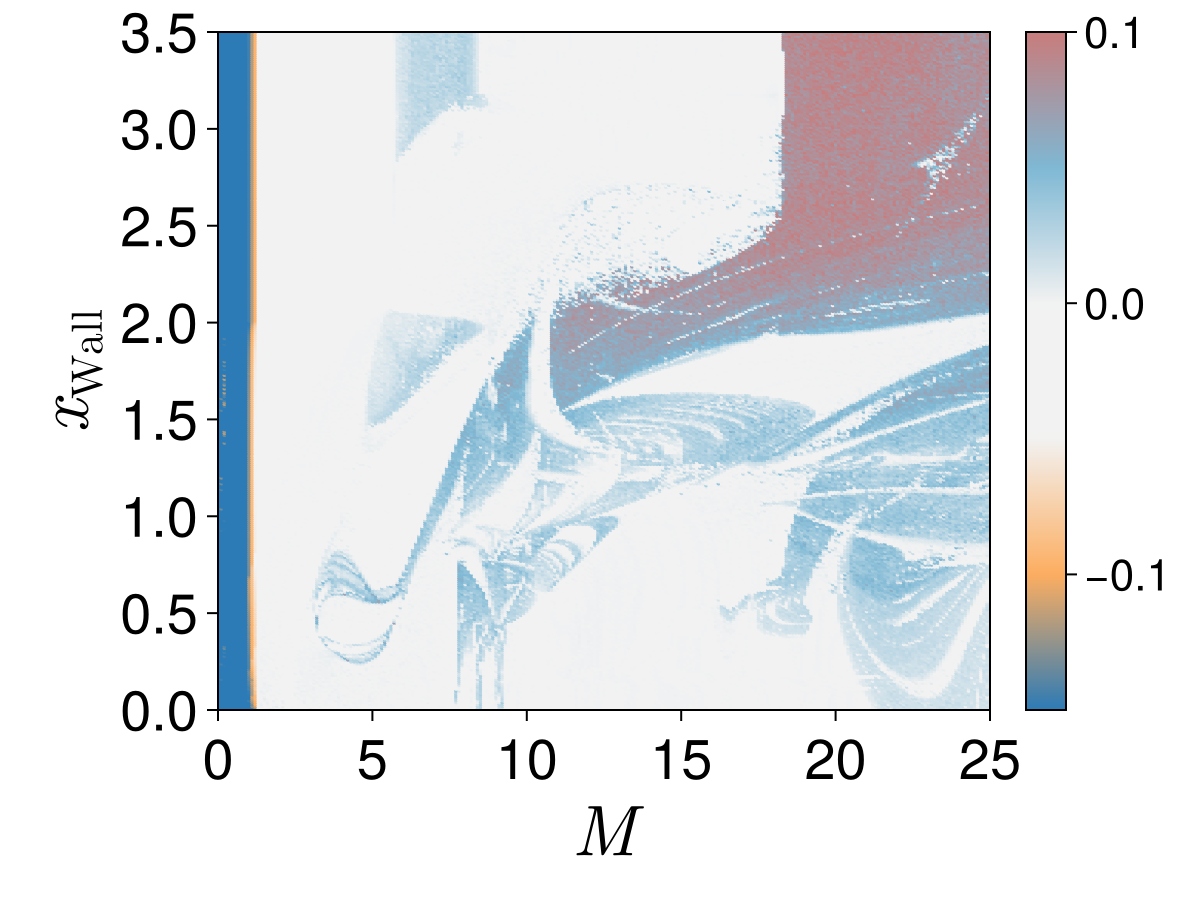}
    \includegraphics[width=0.31\textwidth]{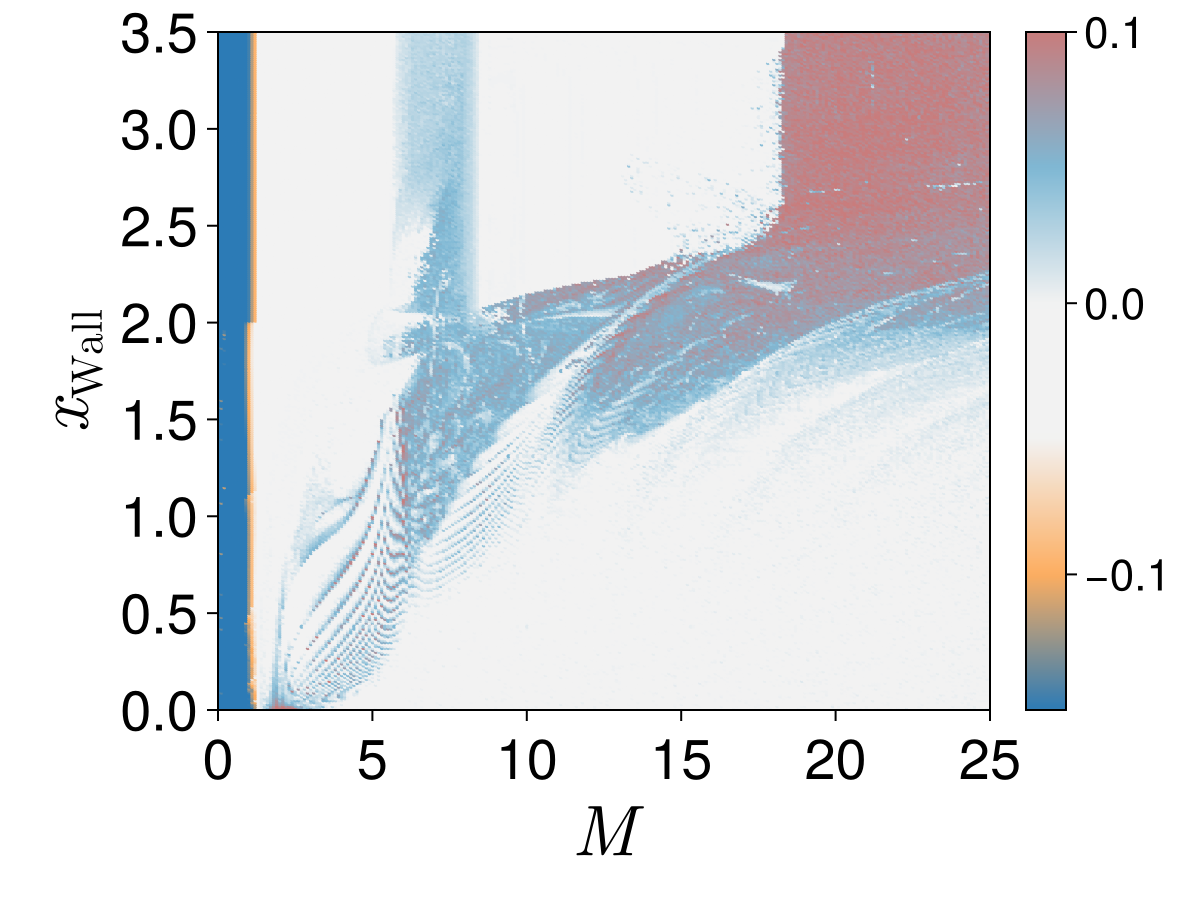}\\
    {\footnotesize (a)\hspace{5cm}(b)\hspace{5cm}(c)}\\[2mm]
    \caption{
    Dynamics in the $x_{\text{wall}}$--$M$ parameter space for different values of the stiffness parameter $A$, where the color denotes the maximum Lyapunov exponent (MLE). Three values of the stiffness parameter $A$ are considered:  
    (a) $A = 0$, (b) $A = 5$, and (c) $A = 100$.  
    The color scale indicates the dynamical regime: dark blue and yellow correspond to negative MLE (stable points), white to near-zero MLE (periodic/quasiperiodic dynamics), and light blue and red to positive MLE (chaotic motion). Other parameters are fixed to $k = 1$, $R = 1.0$, and initial conditions are fixed to $(x_d(0),X(0),Y(0),Z(0)) = (1.0,0.1,0.0,0.0)$.}
    \label{fig: WM}
\end{figure*}

To further understand the dynamical behavior of WPE and the associated transitions in the parameter space, we present results of numerical simulations in the $R$--$M$ parameter space as shown in Fig.~\ref{fig: RM}(a)--(c) for three different values of stiffness parameter $A = 0$, $5$, and $100$, respectively, while keeping fixed $R = 1.0$, $x_\text{wall} = 1.0$ and $k = 1$. We numerically solve the dynamical system in Eq.~\eqref{Eq: Lorenz eq_impact} that describes the WPE dynamics in the non-smooth potential. This numerical solution is then used to calculate the maximal Lyapunov exponent (MLE), which is plotted as color in the parameter space shown in Fig.~\ref{fig: RM} (see Appendix~\ref{App: num details} for details). The black dashed curve is the analytically calculated linear stability boundary for Hopf bifurcations calculated from Eq.~\eqref{eq: stability_curve}. Below this curve, MLE $< 0$ and the WPE motion converges to the stationary equilibrium point located at $(x_d, X, Y, Z) = (0,0,0,0)$ as shown by blue and yellow regions. Above the black dashed curve, a limit cycle emerges, corresponding to MLE $= 0$, as indicated by the gray regions. For larger values of $R$ or $M$, the system exhibits chaotic motion as shown by red regions where MLE $> 0$. We now turn to understand the different transitions in the WPE motion in this parameter space and how they are affected by the presence of the non-smooth potential.

For $A = 0$ (see Fig.~\ref{fig: RM}(a)), corresponding to WPE motion in a smooth simple harmonic potential, a broad red region appears above the stability boundary, indicating that chaotic dynamics dominate at sufficiently large $M$ and $R$ in the absence of any non-smooth effects. Earlier work on walker droplets in a harmonic potential \cite{walker_chaos,durey2018dynamics,perrard2018transition} also demonstrated chaotic dynamics in the high-memory regime. Introducing a piecewise-smooth potential with a small stiffness factor $A = 5$ (see Fig.~\ref{fig: RM}(b)) generates substantial gray patches of periodic motion within the zone that was previously red, i.e., chaotic for $A = 0$. These regions reflect the emergence of periodic dynamics, suggesting that soft impacts suppress chaos and promote regular motion. Increasing the stiffness further to $A = 100$ (see Fig.~\ref{fig: RM}(c)) leads to a parameter space largely dominated by periodic behavior, with only narrow red bands of chaos remaining. Thus, for the parameter ranges considered, increasing the stiffness parameter $A$ progressively suppresses regions of chaotic dynamics in the $R$--$M$ parameter space. However, we note that for large $A$, the narrow chaotic bands extend toward the vicinity of the linear stability boundary (black dashed curve), implying that chaos can arise at smaller values of $R$ and $M$ when the stiffness of the non-smooth potential is high.

\begin{figure}
    \centering
    \includegraphics[width=0.8\linewidth]{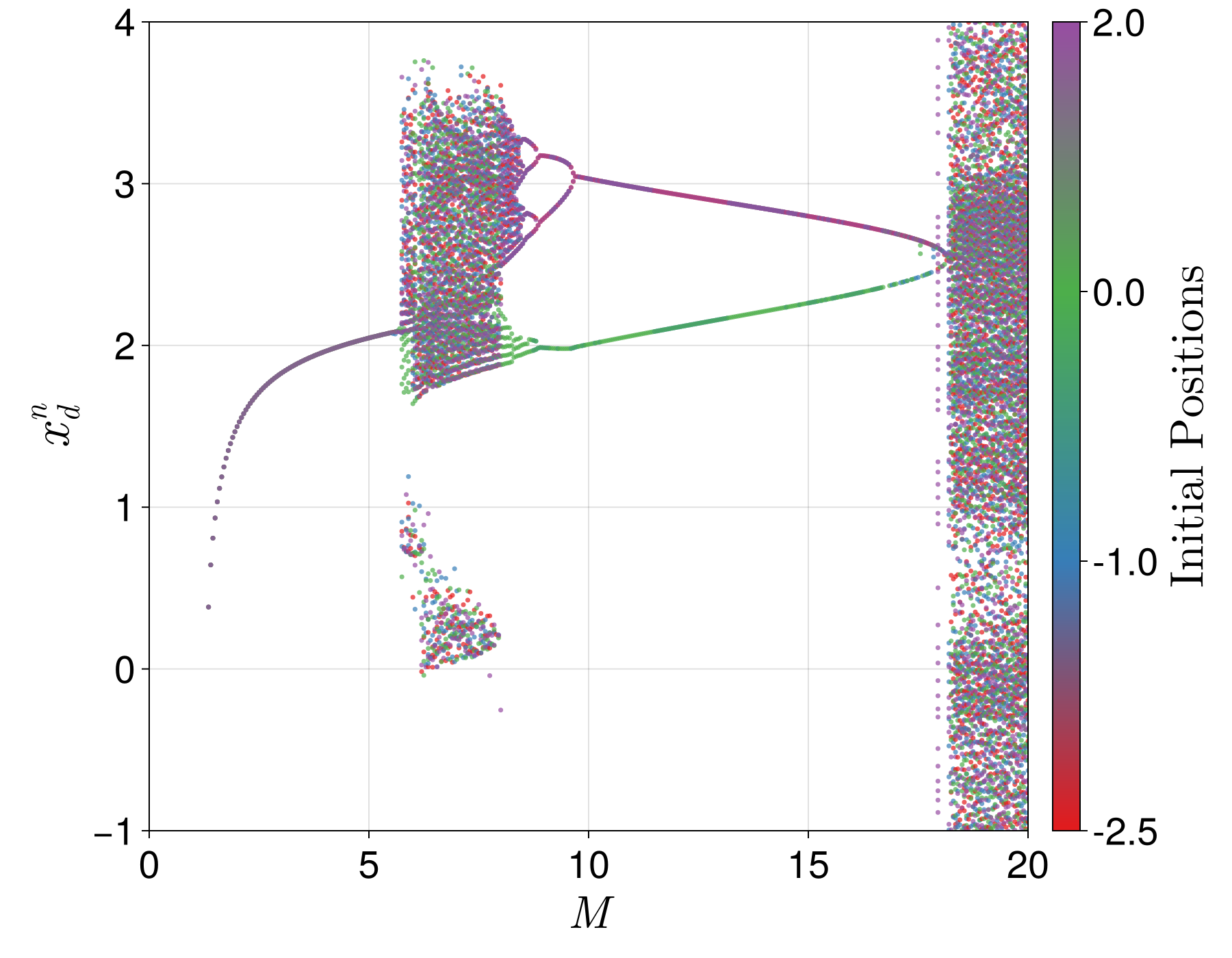}
    \caption{Bifurcation diagram illustrating multistability as a function of the memory parameter $M$ for $x_{\text{wall}} = 1.0$, $R = 1.0$, $k = 1$, $A = 0$ computed over initial droplet positions $x_d(0) \in [-2.5,\, 2.0]$ and initial velocity $X = 1.0$, initial wave memory force $Y = 0.0$, initial wave-field height $Z = 0.0$. The bifurcation diagram is constructed by treating $M$ as the control parameter and recording the particle position $x^n_d$ at the Poincaré section defined by $X = 0$.}
    \label{fig: multi_bif}
\end{figure}

\begin{figure*}[hbtp!]
    \centering
    \includegraphics[width=0.4\linewidth]{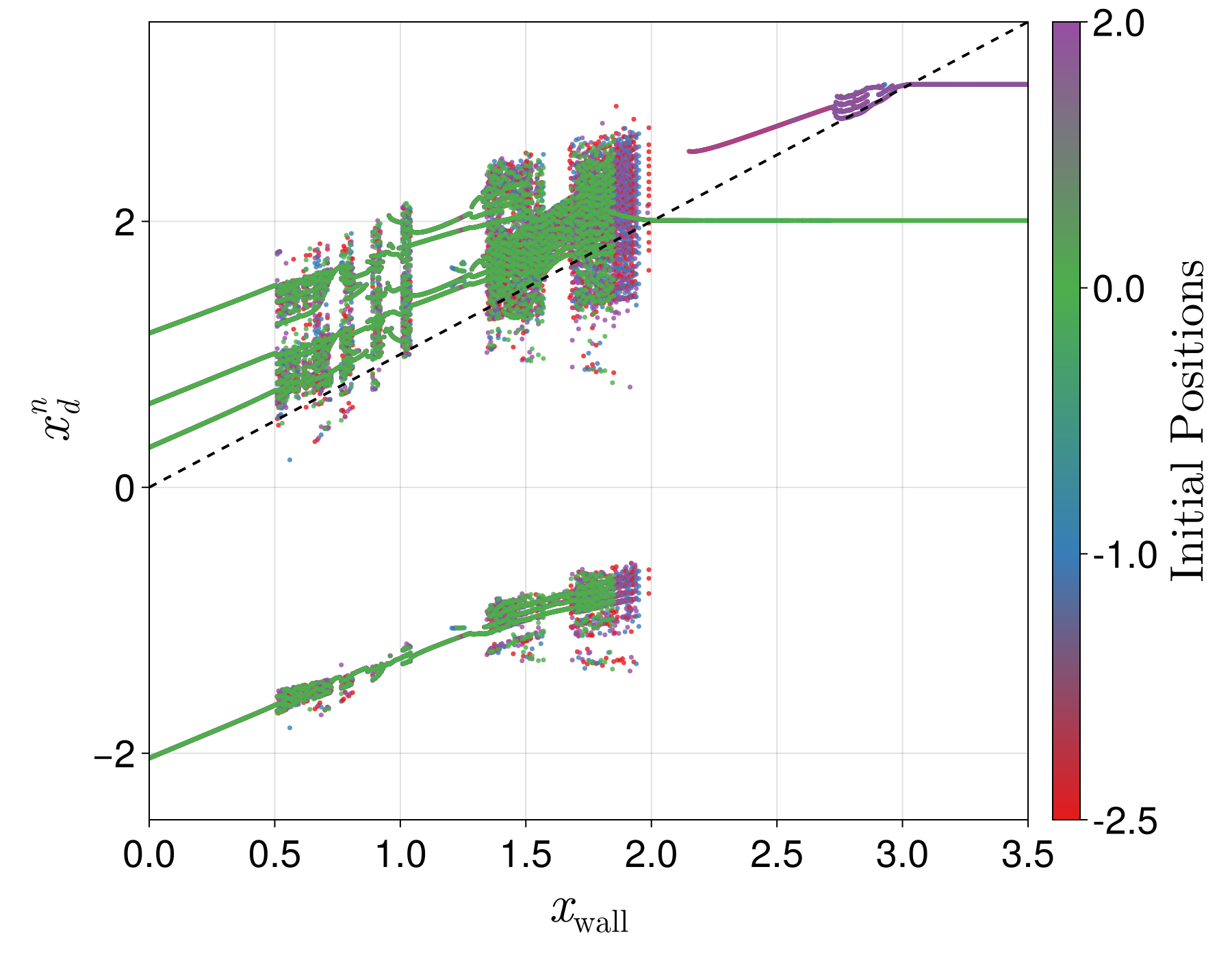}
    \includegraphics[width=0.4\linewidth]{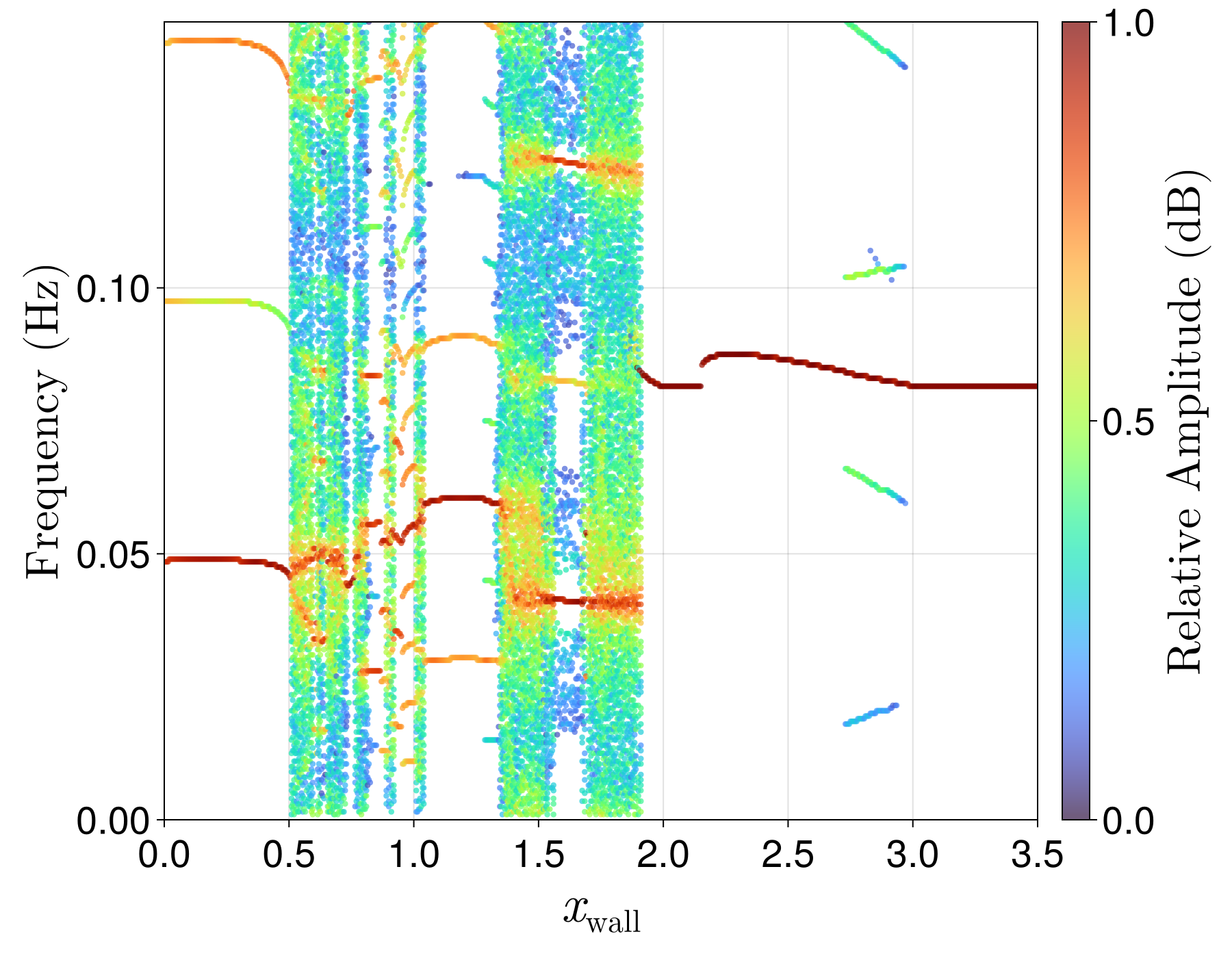}
    \\[-0.3em]
    {\centering\footnotesize (a)\hspace{7cm}(b)\\[3mm]}
    \includegraphics[width=0.4\linewidth]{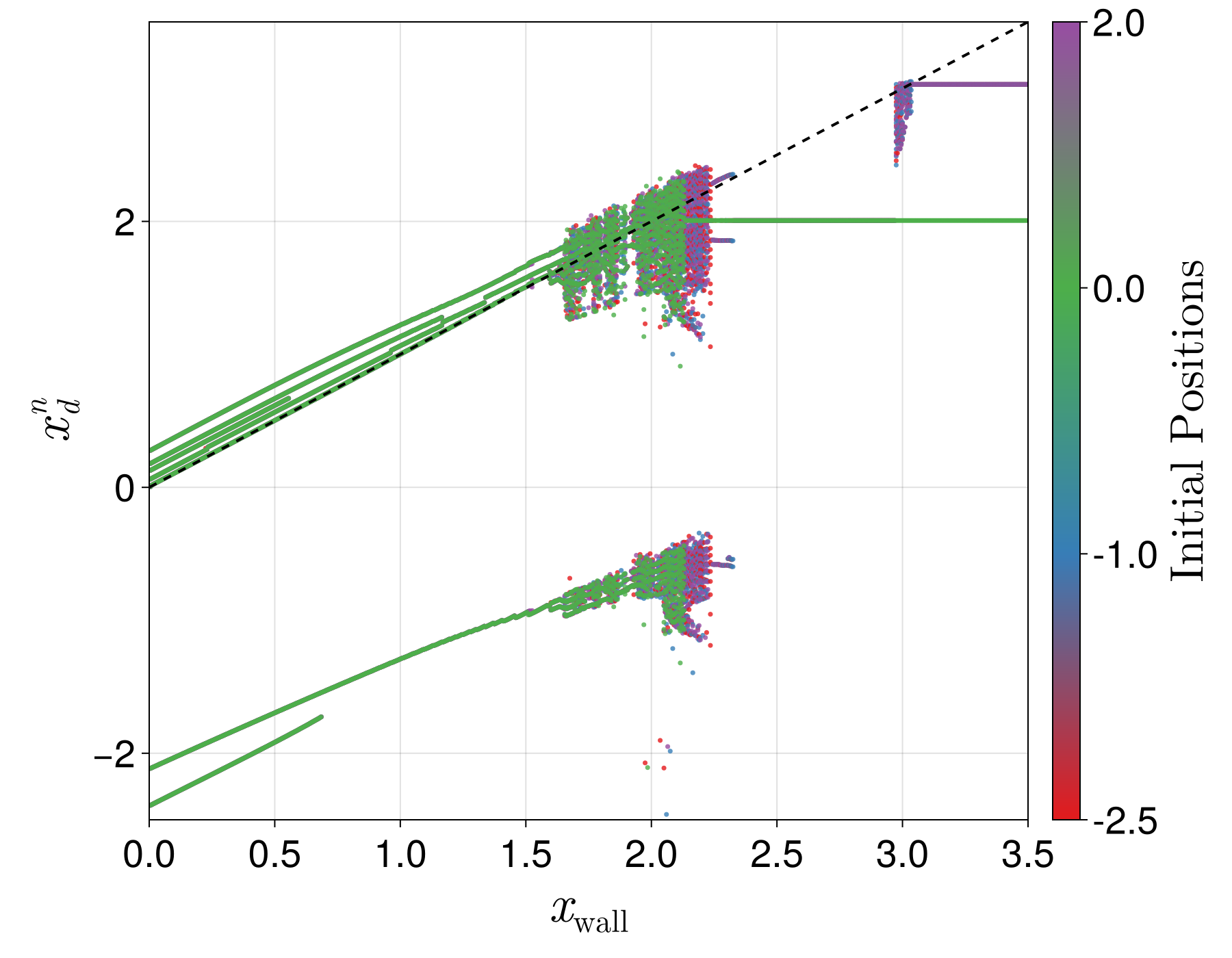}
    \includegraphics[width=0.4\linewidth]{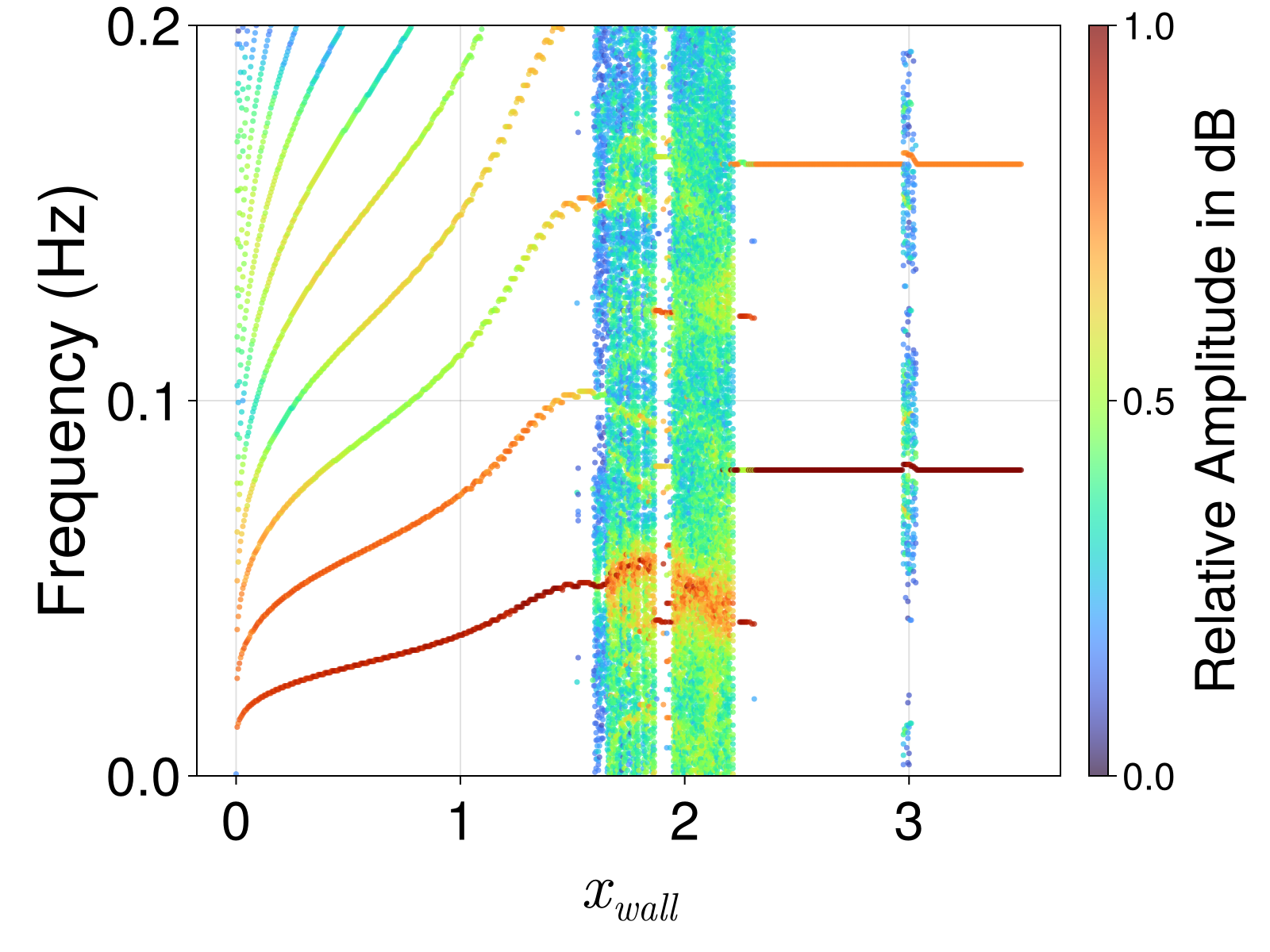}
    \\[-0.3em]
    {\centering\footnotesize (c)\hspace{7cm}(d)\\[3mm]}
    \caption{Bifurcation diagrams and corresponding spectral bifurcation diagrams of the WPE dynamics as a function of the wall position $x_{\text{wall}}$ for $M = 10$ and $R = 1.0$. Panels (a,b) show the state-space and spectral bifurcation diagrams for the soft-wall case $A = 5$, while panels (c,d) correspond to the stiff-wall case $A = 100$. The spectral bifurcation diagrams reveal the transitions between periodic, quasiperiodic, and chaotic dynamics induced by grazing interactions with the wall.
}
\label{fig: bif_5_100}
\end{figure*}

\subsection{Grazing-induced and impact-induced bifurcations}

We now investigate the $x_{\text{wall}}$--$M$ parameter space fixing $R = 1.0$, $k = 1.0$ while varying the stiffness parameter $A = 0$, $5$, and $100$, as shown in Fig.~\ref{fig: WM}(a)--(c), respectively. For all three cases, when the memory parameter $M$ is very small, the WPE dynamics converges to the stationary point at $x_d = 0$ for any wall position $x_\text{wall} > 0.0$, which is evident from the blue/yellow regions corresponding to negative Lyapunov exponents in Figs.~\ref{fig: WM}(a)--(c).

We first consider the no-wall case ($A = 0$). Since no impact occurs in this limit, the dynamics depend only on the memory parameter $M$, and thus the $x_{\text{wall}}$--$M$ parameter space exhibits variation solely along the $M$-axis. As $M$ increases, the system undergoes a Hopf bifurcation, destabilizing the stationary equilibrium and giving rise to periodic oscillations with an MLE equal to zero (white region). Upon further increasing $M$, a narrow vertical band of weakly chaotic dynamics emerges around $M \approx 6$--$8$, characterized by a small positive MLE (light blue). This chaotic region is followed by a return to periodic motion for $8 < M < 18$. Beyond this interval, the dynamics transforms once again to chaos, with strong sensitivity to initial conditions for $M > 18$ (red regions). Hence, we observe two distinct regimes of chaotic motion: ``weak'' chaos (light blue) for $M \approx 6$--$8$ and ``strong'' chaos (red) for $M > 18$.

To further explore these transitions and the possibility of multistability, we compute a bifurcation diagram as shown in Fig.~\ref{fig: multi_bif}, plotting the WPE position $x^n_d$ when its velocity is zero as a function of $M$ for $x_{\text{wall}} = 1.0$, $R = 1.0$. The diagram is constructed using a range of initial particle positions $x_d(0) \in [-2.5,\,2.0]$, while keeping the other initial conditions fixed. From the bifurcation diagram, we observe that the first transition from periodic to chaotic motion near $M \approx 6$ appears to arise through an interior crisis, in which the periodic orbit suddenly gives way to a ``weak'' chaotic attractor. The subsequent return from this chaotic state to periodic motion near $M \approx 8$ occurs via a reverse period-doubling sequence. Within this periodic regime, the system also exhibits multistability: two distinct periodic attractors coexist over the interval $10 \lesssim M \lesssim 18$, each selected by different initial conditions. As $M$ increases further, these two periodic branches approach each other and eventually collide near $M \approx 18$, triggering another transition to ``strong'' chaos. 

We now consider the non-smooth potential with a non-zero stiffness parameter, shown in Figs.~\ref{fig: WM}(b)--(c), where the WPE dynamics vary as a function of both $x_{\text{wall}}$ and $M$. For the softer wall with $A = 5$, the $x_{\text{wall}}$--$M$ parameter space displays a mixture of periodic and chaotic regions, in parameter regimes that were dominated by a single type of behavior for the $A = 0$ case. Two complementary trends emerge: (i) suppression of chaotic regions that were present for $A = 0$, which are replaced by periodic particle motion, and (ii) enhancement of chaos in regimes that were previously purely periodic. Trend (i) is evident for $M \approx 6$--$8$, where the weak chaotic band found in the $A = 0$ case is largely replaced by regular oscillations. In addition, we observe suppression of chaos for $M \gtrsim 18$ and small $x_{\text{wall}} \lesssim 2.0$, as well as intricate interleaving of periodic and chaotic bands. Trend (ii) is observed for $4 \lesssim M \lesssim 18$, particularly at small wall positions, where impacts occur more frequently. In this regime, alternating chaotic bands emerge in place of the fully periodic region seen in the $A = 0$ case.

For the stiffer wall with $A = 100$, several new qualitative features appear. First, chaotic dynamics emerge for small $M$ and small $x_{\text{wall}}$. Second, in the regime $M \approx 6$--$8$, where weak chaos was present for $A = 0$, chaos is now suppressed only when the wall is close to the particle (small $x_{\text{wall}}$), while for larger $x_\text{wall}$ the particle still undergoes chaotic motion. Finally, for large $M$ and $x_{\text{wall}} \lesssim 2.0$, we observe a complete suppression of chaos in favor of robust periodic orbits. In this regime, high memory drives strong self-propulsion, but the stiff wall repeatedly redirects the particle along predictable rebound paths, effectively regularizing the motion. In both the softer and the stiffer wall cases, the chaotic regions exhibit a fine structure consisting of alternating patches of weak chaos (light blue) and strong chaos (red).

A spectral bifurcation diagram (SBD) represents the evolution of the dominant frequencies of a dynamical system as a control parameter is varied, thereby providing a frequency-domain counterpart to the conventional bifurcation diagram. Unlike state-space bifurcation plots, SBDs enable a direct identification of periodic, quasiperiodic, and chaotic dynamics through their characteristic spectral signatures \cite{Spectral@Guha}.

We now study the bifurcation diagram together with the spectral bifurcation diagram of the WPE dynamics as a function of the wall position $x_{\text{wall}}$, for the wall stiffness parameters $A = 5$ and $A = 100$, while adjusting the memory parameter at $M = 10$. This value of $M$ corresponds to a regime in which two coexisting limit cycles are present in the unconfined case ($A = 0$).

For a soft wall with $A = 5$, Fig.~\ref{fig: bif_5_100}(a) shows the bifurcation diagram obtained using a range of initial WPE positions $x_d(0) \in [-2.5,\,2.0]$. At the right end of the diagram, where the wall is sufficiently far from $x = 0$ and impacts do not occur, two distinct coexisting limit cycles are clearly observed, arising from different initial conditions.

The corresponding spectral bifurcation diagram, calculated for the same parameter range using the initial condition $x_d(0) = 2.0$, is shown in Fig.~\ref{fig: bif_5_100}(b). In this region, the spectrum consists of a single dominant frequency along with its harmonics, confirming the purely periodic nature of the motion.


As the wall moves further to $x_{\text{wall}} \approx 3.0$, one of the limit cycles undergoes a grazing interaction with the wall. This grazing event induces a brief transition to quasiperiodic dynamics over a very narrow interval of wall positions. This transition is characterized by a zero Lyapunov exponent and the appearance of two dominant incommensurate frequencies in the spectral bifurcation diagram, as shown in Fig.~\ref{fig: bif_5_100}(b). Upon further decrease in $x_{\text{wall}}$, the trajectory returns to a period-one oscillation, while the second periodic branch remains unaffected in this parameter range.

A second grazing event occurs near after $x_{\text{wall}} \approx 2.0$, now involving the other limit cycle. Although the periodic orbit corresponding to the initial condition $x_d(0) = 0.0$ persists slightly beyond this point, the system transitions to chaotic dynamics for other initial conditions following this second grazing interaction. The onset of chaos is clearly revealed in the spectral bifurcation diagram by the emergence of a continuous broadband frequency spectrum, which is a hallmark of chaotic motion. Notably, within this broadband spectrum, a pronounced spectral component appears at half the frequency of the original period-one orbit. This feature indicates the occurrence of a period-doubling bifurcation prior to the abrupt transition to chaos, consistent with known spectral signatures of chaotic attractors~\cite{Spectral@Guha}.

The abrupt emergence of chaos after grazing suggests that the transition is mediated by a grazing-induced crisis. At $x_{\text{wall}} \approx 1.8$, the remaining periodic orbit corresponding to $x_d(0) = 0.0$ merges into the chaotic band, leading to the loss of multistability. Beyond this point, the system exhibits predominantly chaotic dynamics interspersed with narrow periodic windows. Finally, as the wall moves further inwards to $x_{\text{wall}} \approx 0.5$, the dynamics settle into a periodic orbit, as indicated by the one dominant spectral peak and its harmonic in Fig.~\ref{fig: bif_5_100}(b).

We next examine the bifurcation diagram shown in Fig.~\ref{fig: bif_5_100}(c) for $A = 100$, i.e., when the wall is stiff. Although the overall structure of the dynamics is qualitatively similar to the $A = 5$ case, several notable differences arise in the vicinity of grazing events. When one of the two coexisting periodic trajectories grazes the wall near $x_{\text{wall}} \approx 3.0$, a narrow chaotic band emerges. This behavior is characteristic of narrow-band chaos associated with classical grazing bifurcations~\cite{nordmakmap2020, kundu2012singularities, Banerjee2009}.

The presence of narrow-band chaos is further confirmed by the spectral bifurcation diagram shown in Fig.~\ref{fig: bif_5_100}(d) for the initial condition $x_d(0) = 2.0$, where a continuous frequency band appears over a small parameter interval. As the wall position decreases further, this periodic branch disappears, and all initial conditions converge onto a single remaining periodic attractor.

Upon further inward motion of the wall to $x_{\text{wall}} \approx 2.4$, this surviving limit cycle loses stability and changes to a chaotic attractor. Although the precise mechanism of this transition is not immediately apparent from the state-space bifurcation diagram alone, the spectral bifurcation diagram provides clear insight. Specifically, the appearance of a dominant spectral component at half the frequency of period-one immediately prior to the onset of chaos indicates a period-doubling followed by a border collision bifurcation, consistent with known spectral criteria~\cite{Spectral@Guha}. Beyond this chaotic interval, the system eventually re-enters a regular periodic regime, as evidenced by the persistent single frequency and its harmonics dominance in the corresponding spectral region.

\begin{figure}[hbtp]
    \centering
    \includegraphics[width=0.45\textwidth]{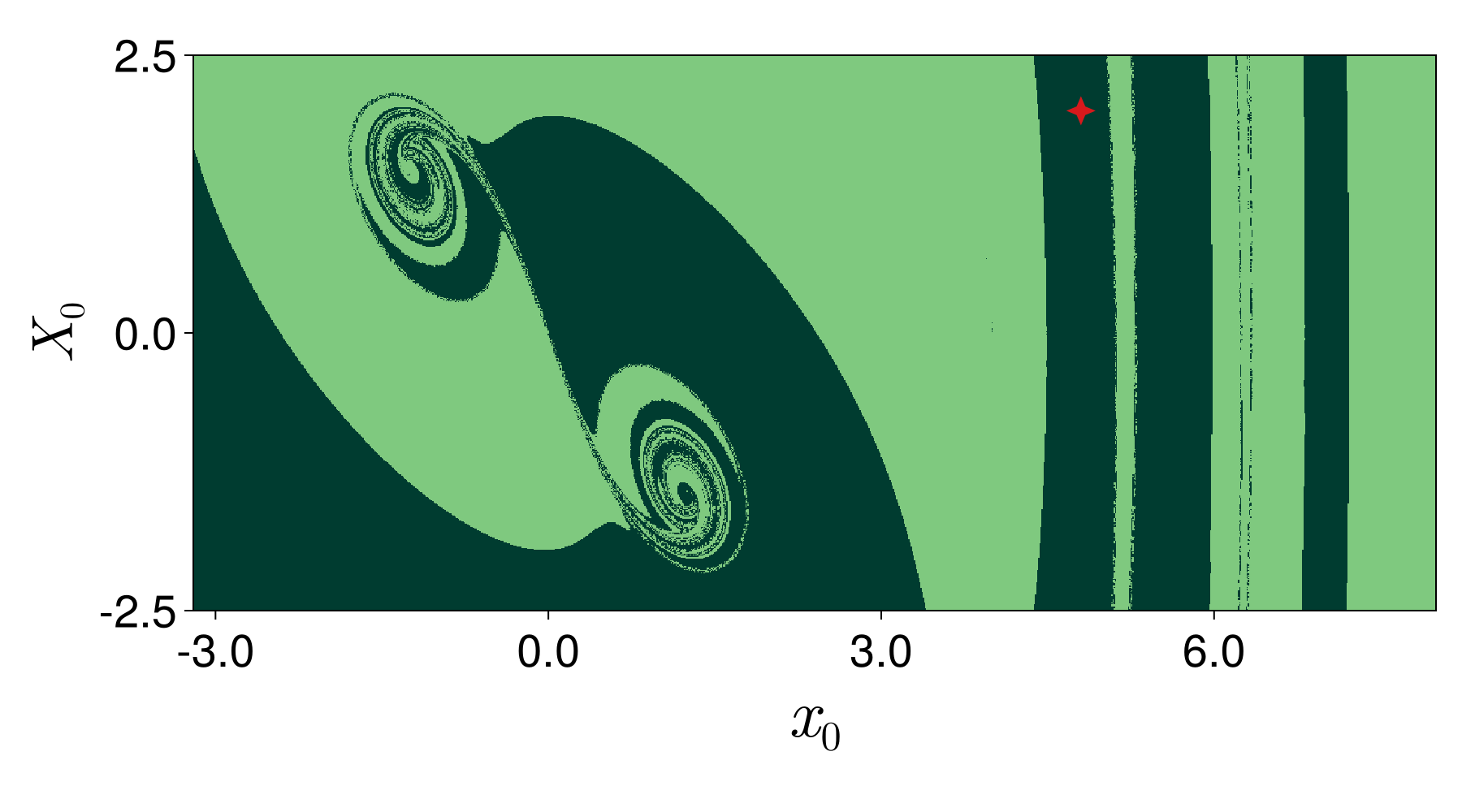}
    \\[-0.3em]
    {\centering \footnotesize (a)\\}
    \includegraphics[width=0.45\textwidth]{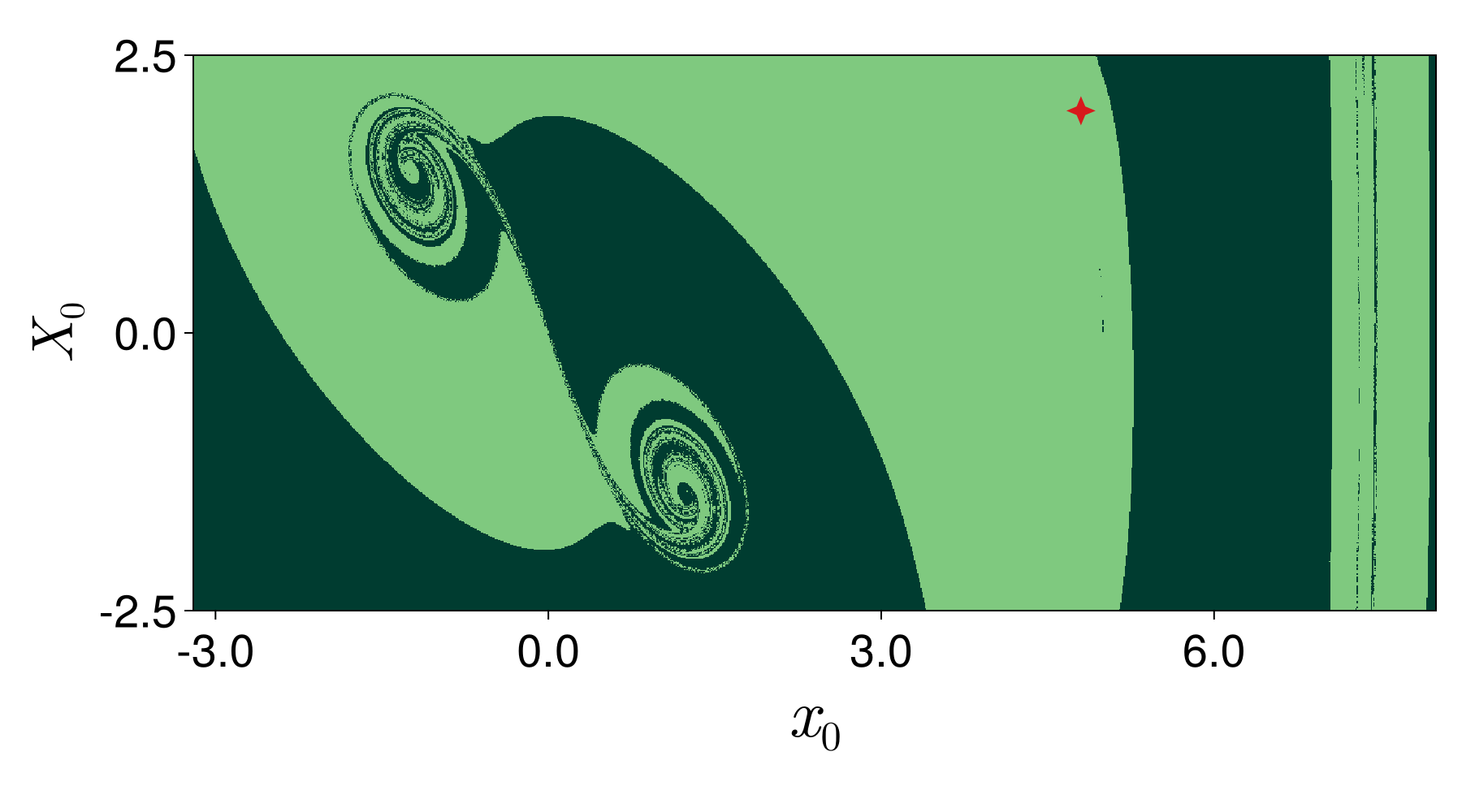}
    \\[-0.3em]
    {\centering \footnotesize (b)\\}
    \includegraphics[width=0.40\textwidth]{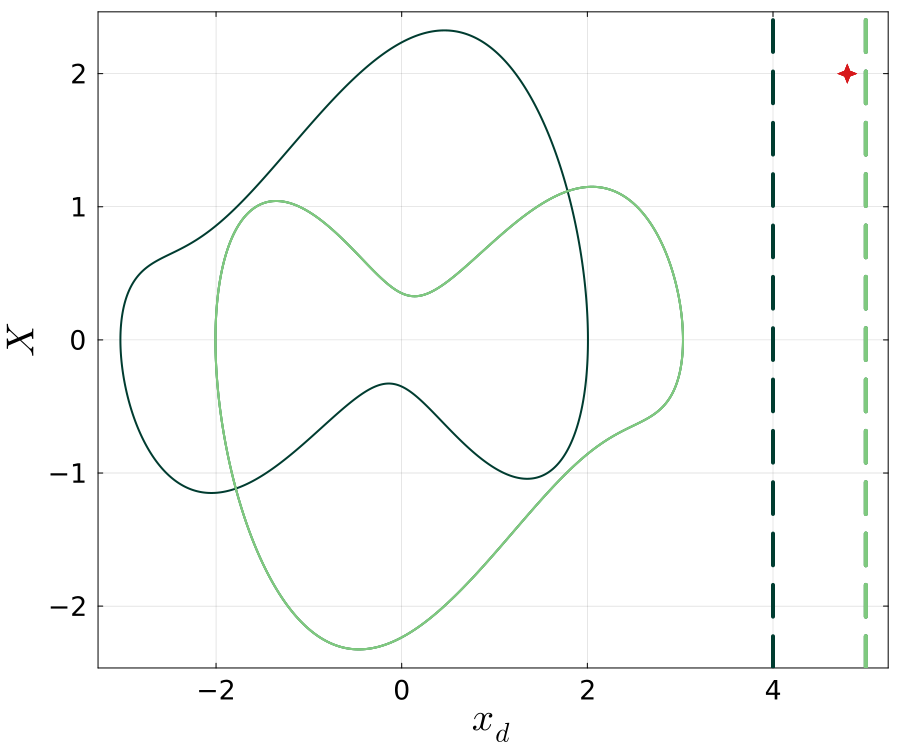}
    \\[-0.3em]
    {\centering \footnotesize (c)\\}
    \caption{Basins of attraction for two coexisting attractors for $A = 50$, $R = 1.0$, $M = 10$, $k = 1$:
    (a) $x_{\text{wall}} = 4.0$ and (b) $x_{\text{wall}} = 5.0$. The red star indicates the initial condition $(4.8,\,2.0,\,0.0,\,0.0)$. (c) Trajectory for the given initial condition with two different wall positions: dark green for $x_{\text{wall}} = 4.0$ and light green for $x_{\text{wall}} = 5.0$. 
    }
    \label{fig: basin}
\end{figure}

\subsection{Fractal basin structure and wall-induced basin switching without impact}

As shown earlier, multistability persists for both $A = 5$ and $A = 100$ when the wall is placed sufficiently far from the origin, i.e., for $x_{\text{wall}} \gtrsim 3.0$. To investigate how different initial conditions converge to different coexisting attractors and how this convergence depends on the wall position, we compute the basins of attraction for the system. The resulting basin structures are shown in Fig.~\ref{fig: basin}. For clarity, we present results for an intermediate wall stiffness $A = 50$. Figure~\ref{fig: basin}(a) corresponds to $x_{\text{wall}} = 4.0$, while Fig.~\ref{fig: basin}(b) shows the basin structure for $x_{\text{wall}} = 5.0$.

For initial conditions in the ranges $x_d(0) \in [-3.0,\,3.0]$ and $X(0) \in [-2.5,\,2.5]$, with $Y(0) = Z(0) = 0.0$, the basin geometry remains largely unchanged between the two wall positions. Nevertheless, within this region the basin exhibits a highly intricate structure: two symmetrically placed lobes appear in phase space, each displaying fine-scale interleaving characteristic of fractal basin boundaries. This fractality implies extreme sensitivity to initial conditions, whereby arbitrarily small perturbations can redirect trajectories from one attractor to the other.

In contrast, for larger initial displacements $x_d(0) \gtrsim 3.0$, the basin structure undergoes a qualitative change as the position of the wall is varied. To illustrate this sensitivity, we highlight a representative initial condition (red star) $(x_d(0),X(0),Y(0),Z(0)) = (4.8,\,2.0,\,0.0,\,0.0).$ For $x_{\text{wall}} = 4.0$, this initial condition lies within the basin of one attractor (dark green), as shown in Fig.~\ref{fig: basin}(a), whereas for $x_{\text{wall}} = 5.0$ the same point belongs to the basin of a different attractor, as seen in Fig.~\ref{fig: basin}(b). The resulting attractor switching, induced solely by changing the wall position, is explicitly illustrated in Fig.~\ref{fig: basin}(c).

In particular, for both location of the walls, the attractors themselves remain well separated from the wall. For a purely structureless classical particle, one would therefore expect the dynamics to reduce effectively to that of a harmonic oscillator. However, in the WPE system the droplet dynamics arises from the coupled evolution of the particle and its associated wave degrees of freedom, leading to intrinsically complex trajectories even in the absence of any boundary.

Importantly, while the wall enters the dynamics only through the nonsmooth force acting on the droplet (via the $\dot X$ equation), its presence nevertheless modifies how these complex wave--particle trajectories explore phase space. Small changes in the wall position alter the regions of phase space where the nonsmooth force becomes relevant, thereby reshaping the basin boundaries and inducing attractor switching. This mechanism does not rely on a direct interaction between the wave field and the wall, but instead on the sensitive interplay between the nonlinear wave-particle dynamics and the nonsmooth potential experienced by the droplet.

This behavior more generally highlights how boundary-induced nonsmooth forces can strongly influence the global organization of phase space, even when the asymptotic attractors themselves remain far from the boundary. 


\section{Discussion and Conclusion}\label{Sec: Conc}



In this work, we introduced an \emph{active soft-impact oscillator} as a minimal model for a one-dimensional walking droplet in a piecewise-smooth potential, coupling a soft-impact nonlinearity to Lorenz-like wave-memory dynamics. The nonlinear interaction between the internal wave degrees of freedom and the non-smooth confining potential generates a rich range of behaviors, including periodic motion, weak and strong chaos, grazing-induced bifurcations, boundary crises, multistability, and invisible attractor switching. Grazing and impact events act as organizing mechanisms that reshape phase-space structure by modifying the coupling between particle motion and wave memory: soft impacts constrain large amplitude excursions, whereas stiff impacts induce abrupt state-space redirections. As a result, both soft and stiff confinement can suppress or promote chaos, producing intricate mosaics of periodic windows and chaotic bands in the \(R\)–\(M\) and \(x_{\text{wall}}\)–\(M\) parameter spaces, and highlighting the sensitivity of active wave--particle dynamics to geometric non-smoothness.

Complementing these state-space observations, spectral bifurcation analysis shows that the associated transitions are accompanied by distinct frequency-domain signatures, including quasiperiodic windows, broadband chaos, and period-doubling routes to chaos. The close correspondence between state-space and spectral diagnostics confirms that grazing interactions not only reshape attractor geometry and basin structure, but also reorganize the underlying frequency content of the dynamics.

Beyond the established phenomenology of classical impact oscillators, our study reveals several dynamical features that have not been reported in the earlier impact oscillator literature. First, the combination of memory-driven forcing and soft impacts produces extended regimes of \emph{weak chaos}, characterized by small but positive Lyapunov exponents, which are absent in traditional systems where chaos typically emerges abruptly at grazing. Second, the interplay between the Lorenz-type WPE dynamics and the non-smooth confinement gives rise to \emph{invisible attractor–switching} i.e. neither attractor physically grazes the wall but variations in the wall position reshape the stable and unstable manifolds in a way that silently redirects entire families of initial conditions from one attractor to another. This mechanism is different from classical invisible grazing or dangerous bifurcations~\citep{Banerjee2009}. Third, we identify parameter regions in which impacts \emph{stabilize} the dynamics i.e. suppressing chaos and producing wide periodic windows. Such reversible chaos–order transitions, controlled solely by wall compliance and memory, have not been previously demonstrated in impact oscillators.

The relevance of these results to walking droplet experiments is twofold. The nonlinear behaviors identified here including grazing-induced bifurcations, crisis-induced chaos, multistability, and fractal basin boundaries provide concrete, testable predictions for droplets interacting with non-smooth potentials. Moreover, piecewise-smooth confinement of the type considered here is experimentally plausible. Harmonic trapping of walking droplets has previously been realized using spatially varying magnetic fields acting on ferrofluid-coated droplets~\citep{Perrard2014a}. Superimposing a steep localized magnetic gradient onto such a harmonic field would create a controllable soft-impact region, enabling direct experimental tests of the predicted transitions between periodic, multistable, and chaotic dynamics.

Beyond walking droplets, these results connect naturally to the broader framework of hydrodynamic quantum analogs in generalized pilot-wave dynamics. Our study extends this paradigm to non-smooth potentials, a setting central to quantum impact systems. The piecewise-quadratic potential analyzed here closely parallels recent studies of quantum soft-impact oscillators~\citep{Mukherjee2025SoftImpactQLE, Acharya2025QuantumImpactSSRN}, where parameter variation induces transitions between periodic, multiperiodic, and chaotic regimes via grazing events and crises. More broadly, the active soft-impact oscillator introduced here unifies elements of active matter, nonlinear oscillators, impact dynamics, and hydrodynamic quantum analogs within a single framework. It thus provides a versatile platform for exploring how activity, memory, and non-smoothness collectively generate complex nonlinear dynamics, with future directions including experimental realization, extensions to higher dimensions, detailed basin-geometry analysis, and systematic classical--quantum comparisons.

\section*{Acknowledgments}
We thank Dr. Arnab Acharya for fruitful discussions. T Mukherjee acknowledges financial support from
UGC, Govt. of India. R.V. acknowledges the support of the Leverhulme Trust [Grant No. LIP-2020-014]. 

\bibliography{main}
\bibliographystyle{apsrev4-2}

\appendix

\section{Numerical details}\label{App: num details}
{\color{black} Eq.~\eqref{Eq: Lorenz eq_impact} were integrated numerically using the Julia ecosystem, primarily the \texttt{DifferentialEquations.jl} package~\cite{rackauckas2017differential} together with the \texttt{DynamicalSystems.jl} framework~\cite{datseris2018dynamical}. The piecewise-smooth nature of the equations, arising from impacts with the wall, was handled using an event-driven callback that detects the wall-crossing condition with high accuracy and applies the instantaneous velocity reset. For each parameter set, trajectories were computed with a fixed numerical time step $\Delta t = 10^{-3}$ (unless stated otherwise). Depending on the analysis, an initial transient of $T_{\mathrm{tr}} = 2000$ time units was discarded, and the system was integrated for a total duration of $T_{\mathrm{tot}} = 5000$ time units to guarantee convergence to the long-term attractor. The maximal Lyapunov exponent was computed using the implementation in \texttt{DynamicalSystems.jl}~\cite{datseris2018dynamical}, which follows the standard Benettin--Wolf~\cite{benettin1980lyapunov} algorithm based on repeated renormalization of tangent vectors via QR decomposition. This method provides a robust estimate of the exponential divergence rate of nearby trajectories and is well-suited for piecewise-smooth or weakly chaotic systems.}

\end{document}